\documentclass[12pt,preprint]{aastex}

\shorttitle{Chromospheric brightenings during flux emergence}
\shortauthors{Guglielmino et al.}

\begin{document}

\title{HINODE Observations of Chromospheric Brightenings in the\\ 
\ion{Ca}{2} H Line during small-scale Flux Emergence Events}

\author{S. L. Guglielmino and F. Zuccarello}
\affil{Dipartimento di Fisica e Astronomia, Universit\`a di Catania, 
    Catania, I 95123}
\email{salvo.guglielmino@oact.inaf.it}

\author{P. Romano}
\affil{INAF - Osservatorio Astrofisico di Catania, Catania, I 95123}

\and

\author{L. R. Bellot Rubio}
\affil{Instituto de Astrof\'isica de Andaluc\'ia (CSIC), Granada, E 18080}

\begin{abstract}

\ion{Ca}{2} H emission is a well-known indicator of magnetic activity 
in the Sun and other stars. It is also viewed as an important
signature of chromospheric heating. However, the \ion{Ca}{2} H line
has not been used as a diagnostic of magnetic flux emergence from the
solar interior. Here we report on Hinode observations of chromospheric
\ion{Ca}{2} H brightenings associated with a repeated, small-scale
flux emergence event. We describe this process and investigate the
evolution of the magnetic flux, G-band brightness, and \ion{Ca}{2} H
intensity in the emerging region. Our results suggest that  
energy is released in the chromosphere as a consequence of 
interactions between the emerging flux and the pre-existing 
magnetic field, in agreement with recent 3D numerical simulations.

\end{abstract}


\section{Introduction}

Numerical simulations predict that magnetic flux emerging into the
solar atmosphere interact and reconnect with the pre-existing
chromospheric and coronal field. This suggests that flux emergence is
a relevant source of energy for the chromosphere
\citep{Archontis:04,Archontis:05}. The efficiency of the interaction
and the consequent heating seem to depend on the geometry of the two
flux systems
\citep{Galsgaard:07}. While at large scale these results have been
confirmed by high-resolution observations
\citep{Moreno:08,Zuccarello:08}, the role of small-scale emergence
events in the heating of the upper atmospheric layers, as described by
\citet{Isobe:08}, is still lacking observational confirmation.

In the absence of other diagnostics, heating events in the
chromosphere can be detected through the intensity profiles of the
\ion{Ca}{2} H and K lines. The correlation between flux emergence and
\ion{Ca}{2} chromospheric emission was analyzed in detail by
\citet{Balasu:01}, who observed anomalous profiles in the K line in an
emerging active region. However, very few examples of small-scale
transient \ion{Ca}{2} brightenings have been reported in the
literature, and most of them are related to flux cancellation 
events \citep[e.g.][]{Bellot:05}.

The \emph{Hinode} satellite, with its unprecedented spatial
resolution, offers for the first time the possibility to investigate
the processes that occur in the chromosphere during the emergence of
magnetic flux at small spatial scales. In this Letter we analyze 
simultaneous chromospheric and photospheric observations of an
emerging flux region taken with the Solar Optical Telescope aboard
\emph{Hinode}. During the emergence event, strong brightenings were
detected in the \ion{Ca}{2} H line core without relevant
counterparts in G-band intensity, which suggests that chromospheric
heating did occur at the site of flux emergence.

\section{Observations and data reduction}

On 2007 September 30, as part of the \emph{Hinode} Operation Plan 14
(\emph{Hinode}/Canary Islands Campaign), the active region NOAA 10971
was observed by the Solar Optical Telescope \citep[SOT;][]{Tsuneta:08}
onboard \emph{Hinode} \citep{Kosugi:07}. The field of view (FoV) was
centered at solar coordinates ($174\arcsec$, $-79\arcsec$), 
i.e., $11^{\circ}$ away from disk center.

The SOT spectro-polarimeter \citep[SP;][]{Tsuneta:08} performed six
raster scans of the active region from 08:00 to 14:00 UT, acquiring
the Stokes I, Q, U, and V profiles of the photospheric
\ion{Fe}{1} lines at 630.15 nm and 630.25 nm. The FoV covered
by the SP observations is $164\arcsec \times 164\arcsec$, with an
effective pixel size of $0.32\arcsec$ (Fast Map mode). Simultaneously,
the SOT Broadband Filter Imager (BFI) acquired filtergrams in the
core of the \ion{Ca}{2} H line ($396.85 \pm 0.3 \;\textrm{nm}$) and in
the G band ($430.5 \pm 0.8 \;\textrm{nm}$), while the Narrowband
Filter Imager (NFI) obtained shuttered Stokes I and V filtergrams in
the wings of the \ion{Na}{1} D1 line at $589.6 \;\textrm{nm}$. 
The BFI images have a spatial
sampling of $0.05\arcsec$/pixel (G-band) and $0.1\arcsec$/pixel
(\ion{Ca}{2} H), while that of the NFI filtergrams is $0.16\arcsec$/pixel. 
The BFI and NFI time series have a cadence of one minute and extend from 
07:00 to 17:00 UT, with a small gap between 10:05 and 10:20 UT.

We have corrected the SOT/SP and SOT/FG images for dark current, flat
field, and cosmic rays with standard SolarSoft routines. Besides obtaining
photospheric and chromospheric information through corrected G-band
and \ion{Ca}{2} H filtergrams, we have constructed magnetograms from
the \ion{Na}{1} D1 Stokes I and V images acquired $\pm 156$ m\AA{} off
the line center. From the ratio
\begin{displaymath}
\frac{V}{I}=\frac{1}{2}\left( \frac{V_{\rm blue}}{I_{\rm blue}} + \frac{V_{\rm red}}{I_{\rm red}} \right)
\end{displaymath}
we calculate the magnetic flux density $\Phi_{\rm d}$ using the weak
fied approximation \citep{Stix:02} as
\begin{displaymath}
\Phi_{\rm d}= 8 \times 10^3 \: \frac{V}{I} \;\;\; [\textrm{Mx cm}^{-2}] 
\end{displaymath}
\citep[see][]{Guglielmino:08}. To first order, the magnetograms computed in 
this way are not affected by Doppler shifts. We remind the reader
that, at disk center, the \ion{Na}{1} D1 line refers to the upper
photospheric layers and not to the chromosphere.

For each raster scan of the SP, the profile with the minimum total
polarization degree,
$P=\left[\left(Q^{2}+U^{2}+V^{2}\right)/I^{2}\right]^{1/2}$, was
selected as a reference profile. All the spectra in the scan were
normalized to the continuum of this profile and corrected for
limb darkening. Also, a stray light profile was computed by averaging
the reference profiles of the six scans.

Adopting a grid paradigm, we have inverted the spectra with $P > 2\%$
using the SIR code \citep{RuizIniesta:92}. The inversion yields the
temperature stratification in the range $-4.0 < \log \, \tau_{5} < 0$
($\tau_{5}$ is the optical depth of the continuum at $500
\;\textrm{nm}$), together with the magnetic field strength,
inclination and azimuth angles in the line-of-sight (\emph{los})
reference frame, the \emph{los} velocity, and the magnetic filling
factor, assuming these quantities to be constant with height. Azimuth
and inclination angles have been transformed to the local solar frame,
whereas the \emph{los} velocity has been calibrated using the mean
quiet-Sun intensity profile computed from pixels with $P < 0.5\%$,
following the procedure of \citet{Marti:97}.

Finally, all the SOT/FG and SOT/SP images have been aligned through cross-correlation algorithms.

\section{Results}

\begin{figure}
\epsscale{1.}
\plotone{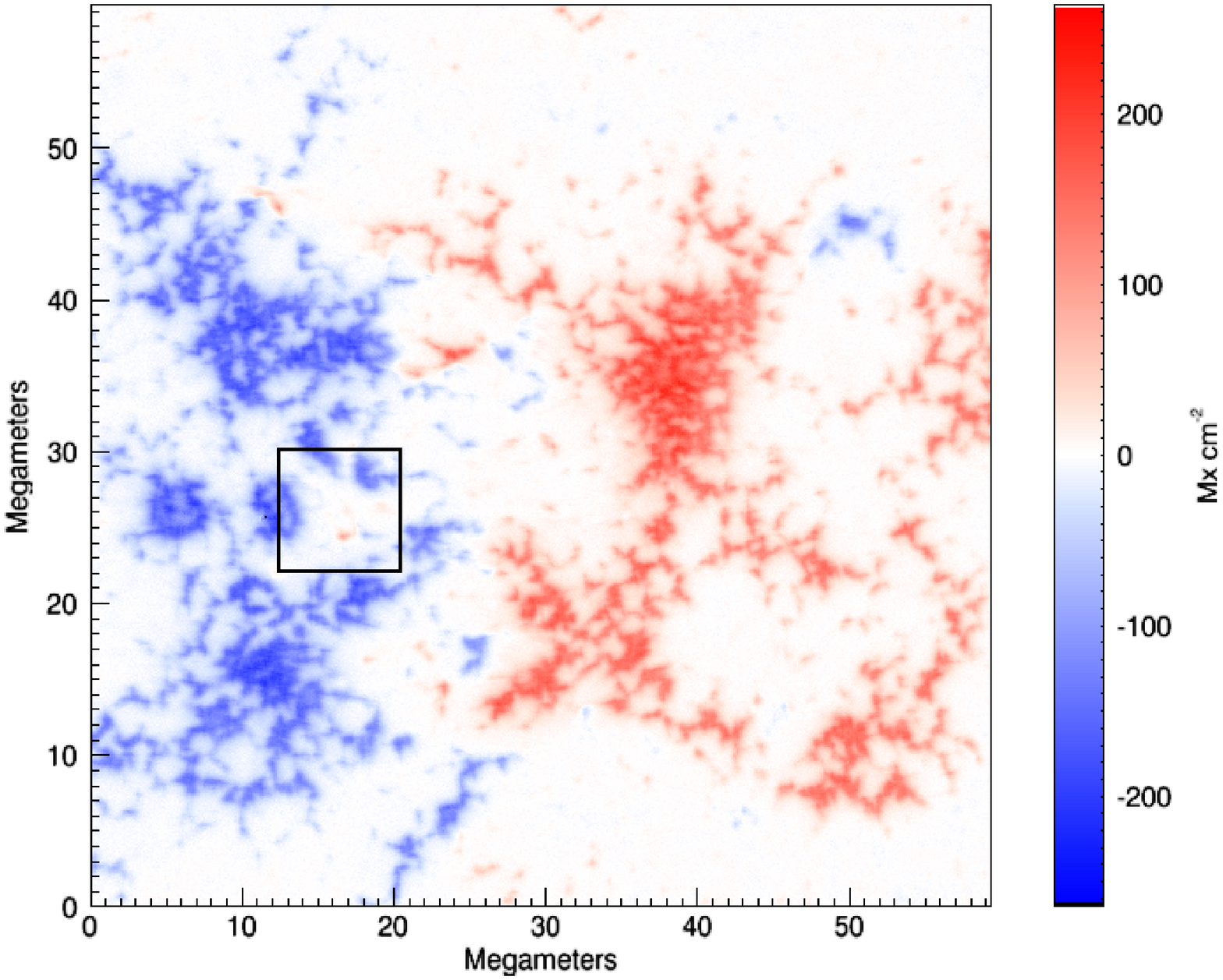}
\caption{Vertical component of the magnetic field obtained from the \ion{Na}{1} magnetograms at 07:50 UT, at the very beginning of the emergence event (FoV $\sim 60 \times 60 \;\textrm{Mm}^{2}$). The square, with a FoV of $8 \times 8 \;\textrm{Mm}^{2}$, indicates the location of the emerging flux region analyzed in this Letter. \label{fig1}}
\end{figure}

\begin{figure}
\epsscale{.7}
\plotone{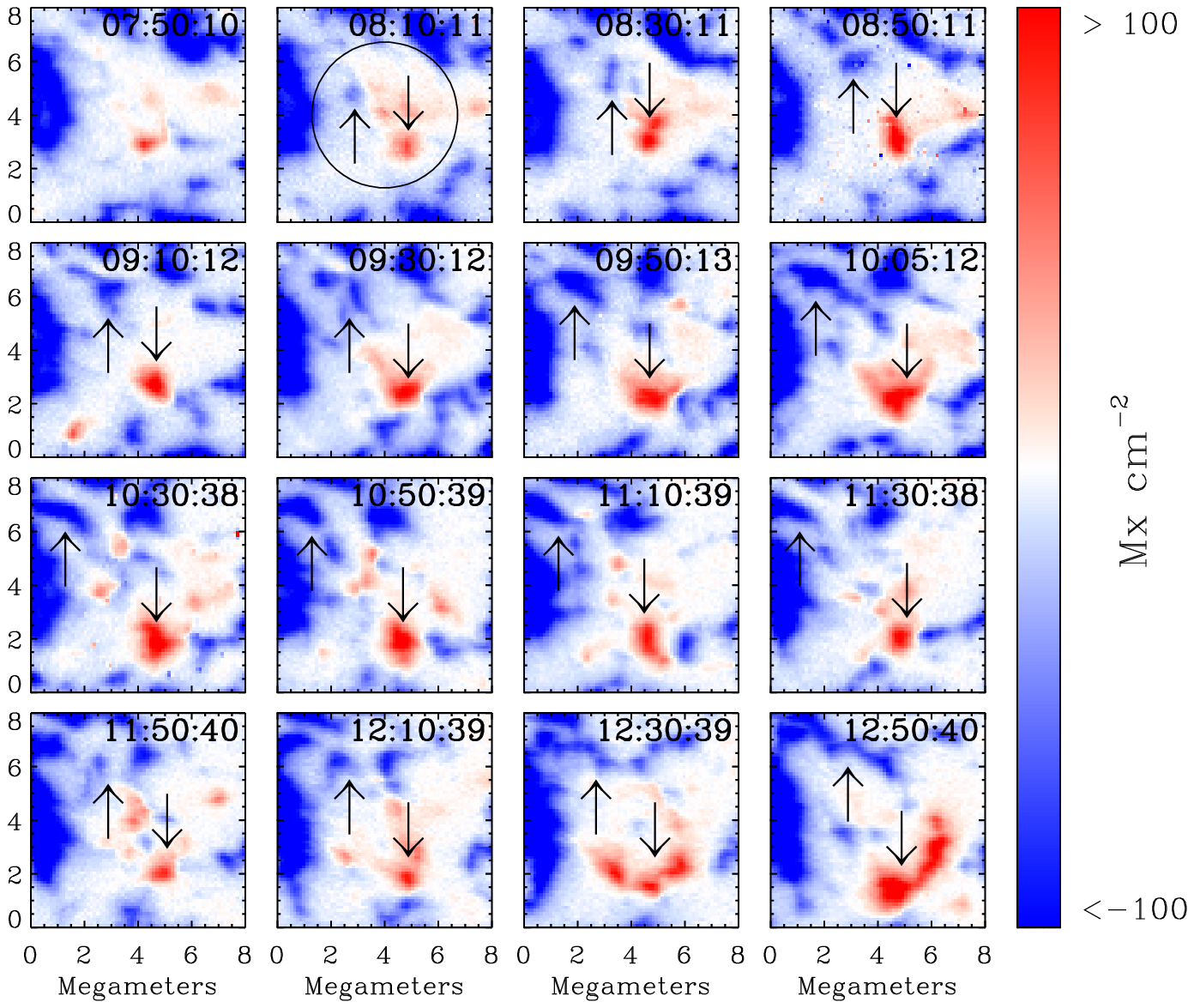}
\plotone{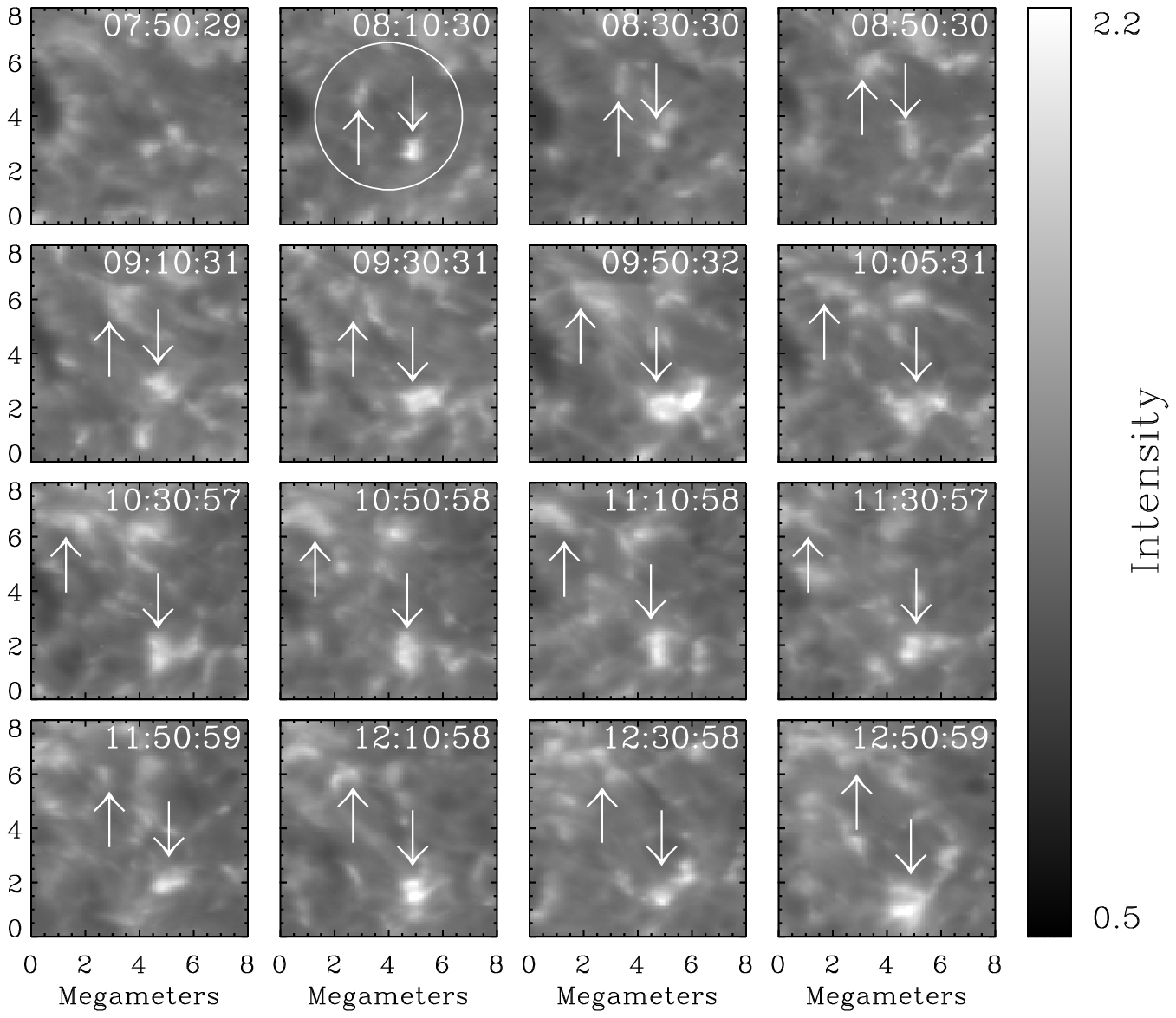}
\caption{\emph{Top}: Temporal sequence of \ion{Na}{1} D1 magnetograms, starting from 
07:50 UT, with a FoV of $8 \times 8 \;\textrm{Mm}^{2}$. Arrows
indicate footpoints of opposite polarity. In the second frame the
circle encloses the region highlighted in
Fig.~\ref{fig4}. \emph{Bottom}: Same, for the \ion{Ca}{2} H
filtergrams. Major Ca intensity enhancements are associated with 
the positive footpoint. Note that the brightening at 09:50 UT occurred above
the contact region between the positive flux element of the emerging
bipole and the negative ambient magnetic field. \label{fig2}}
\end{figure}

\begin{figure}
\epsscale{1.1}
\plottwo{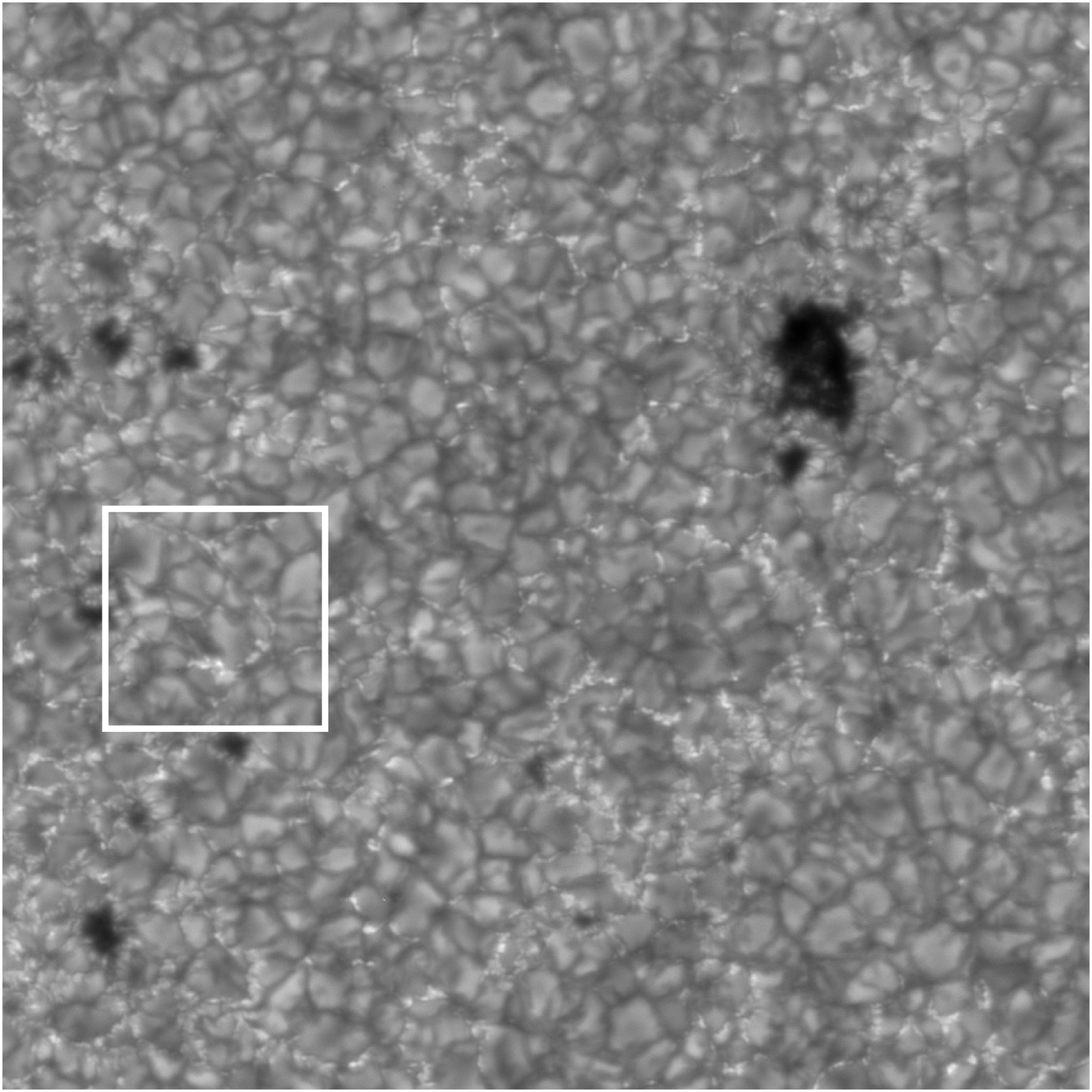}{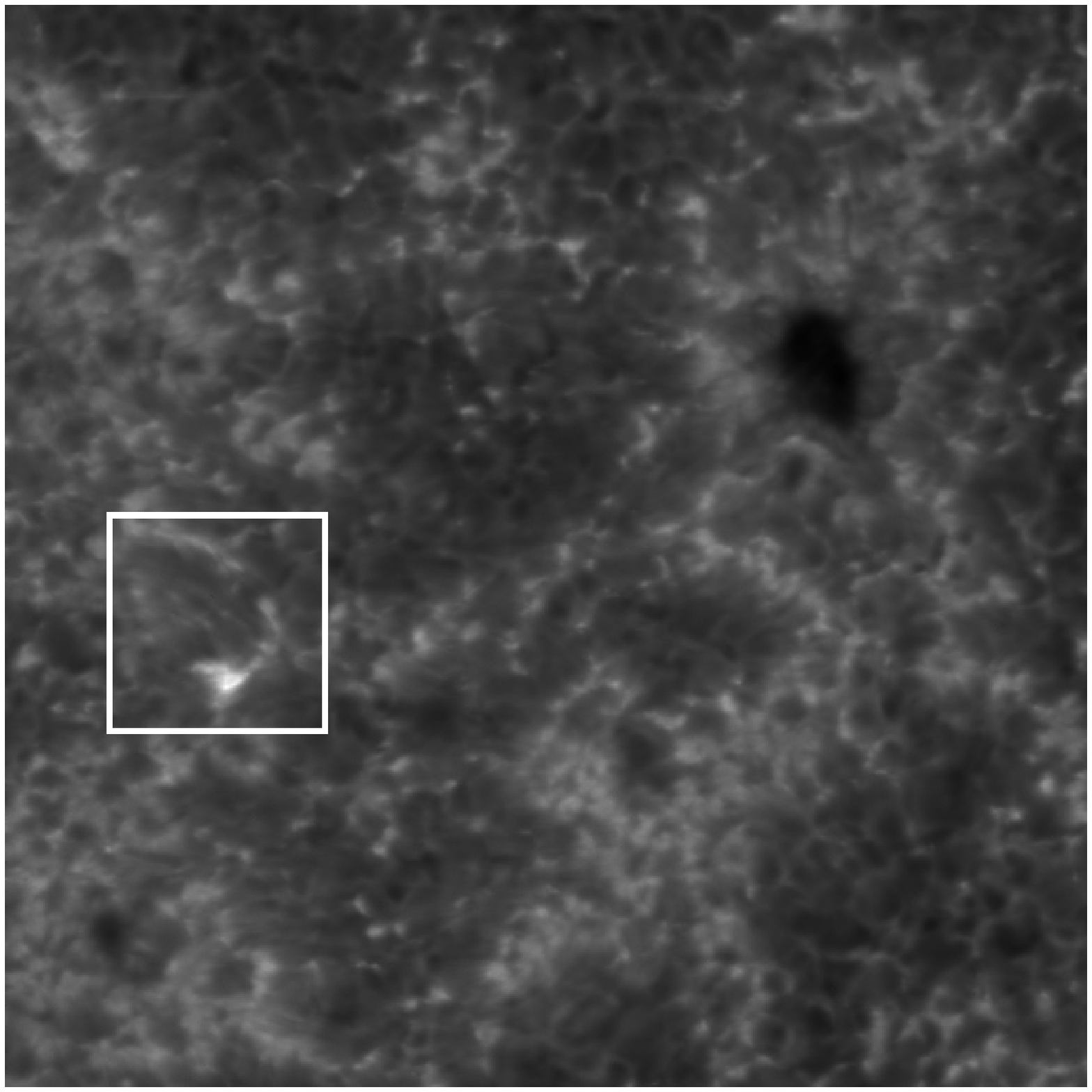}
\caption{G-band (\emph{left}) and \ion{Ca}{2} H (\emph{right}) 
filtergrams (FoV $\sim 80 \times 80 \;\textrm{Mm}^{2}$) acquired 
simultaneously at the time of the second peak in
\ion{Ca}{2} H brightness (12:41 UT). The squares have a FoV of $8
\times 8 \;\textrm{Mm}^{2}$. They contain a bright structure with a
shape similar to that of the emerging region observed with the SOT/SP
at 12:23 UT (see Fig.~\ref{fig4}). \label{fig3}}
\end{figure}

NOAA 10971 has a classical bipolar $\beta$ configuration, as can
be seen in the \ion{Na}{1} D1 magnetogram of Fig.~\ref{fig1}. The
various SOT instruments recorded the emergence of a small
bipolar region, which appeared at the internal edge of the main
negative polarity. Figure~\ref{fig2} displays a temporal sequence of
\ion{Na}{1} D1 magnetograms with a cadence of about 20 minutes for
the $8 \times 8$ Mm$^2$ area marked in Fig.~\ref{fig1}. The first
magnetogram of the sequence, acquired at 07:50 UT, shows the presence
of a positive-polarity knot. In the following magnetograms we clearly
recognize an increase in its area, as well as the appearance of a
negative-polarity patch. The subsequent evolution is characterized by
the separation of the opposite magnetic polarities, as indicated by
the arrows. The corresponding temporal sequence of \ion{Ca}{2} H
filtergrams is also displayed in Fig.~\ref{fig2} and shows
transient brightness enhancements at the location of the positive
footpoint. This emergence event led to the appearance of bright points
in the G band (Fig.~\ref{fig3}, left panel) and to intensity
enhancements in the \ion{Ca}{2} H line core (Fig.~\ref{fig3}, right
panel).

\begin{figure}
\epsscale{1.0}
\plotone{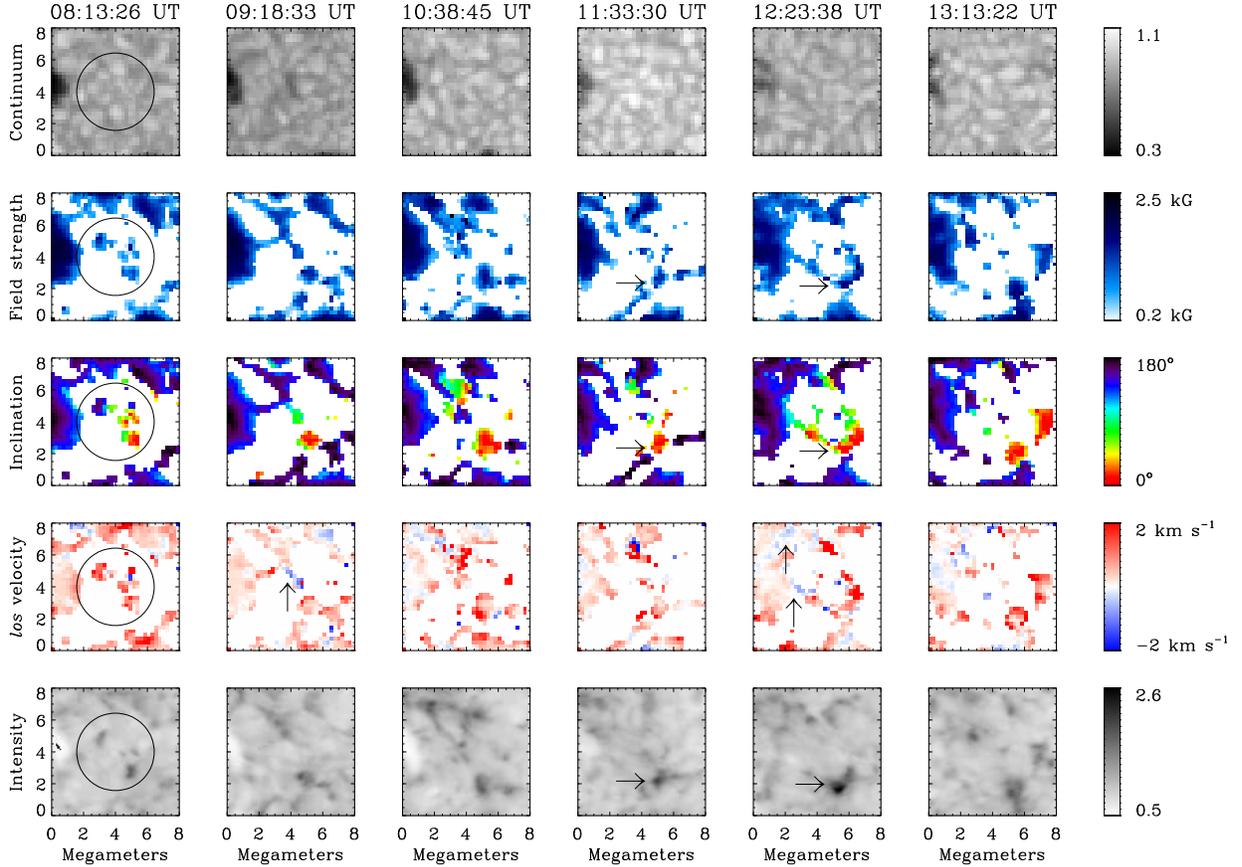}
\caption{Evolution of physical parameters in the emerging 
flux region. From top to bottom: continuum intensity, magnetic field
strength, inclination, and \emph{los} velocity, as derived from the
inversion of the SOT/SP scans. The FoV is $8 \times 8
\;\textrm{Mm}^{2}$, corresponding to the squares marked in
Fig.~\ref{fig1} and~\ref{fig3}. North is up and West to the right. The
white background represents non-inverted pixels. The emerging bipolar
region is enclosed with a circle in the first scan. In the last row 
we show \ion{Ca}{2} H filtergrams, with reversed color scale. 
Vertical arrows indicate blueshifts, whereas horizontal arrows indicate 
the sites of chromospheric brightenings. \label{fig4}}
\end{figure}

Maps of the physical parameters derived from the SP raster scans are
displayed in Fig.~\ref{fig4}. They demonstrate the rapid evolution of
the small bipolar region: the changes in, e.g., the magnetic field
distribution (second and third rows) indicate a very dynamic
phase. The small bipolar region shows an emergence zone, i.e., a
region between the two main polarities with horizontal fields
\citep{Lites:98} in which upflows of $\sim 1 \;\textrm{km s}^{-1}$ 
can be seen at 9:18 UT and 12:23 UT. The footpoints of the emerging
region exhibit vertical fields and downflows of $1.5 - 2 \;\textrm{km
s}^{-1}$. The initial photospheric total flux content of the emerging
region is $1.4 \times 10^{19} \;\textrm{Mx}$, classifying as a small
ephemeral region. The bipole axis was inclined about $45
\degr$ to the north-south direction in the first raster scan, but this
angle varied with time. The negative-polarity footpoint soon merged
with the dominant negative flux of the active region, disappearing as
an individual feature.

\begin{figure}
\epsscale{.93}
\plotone{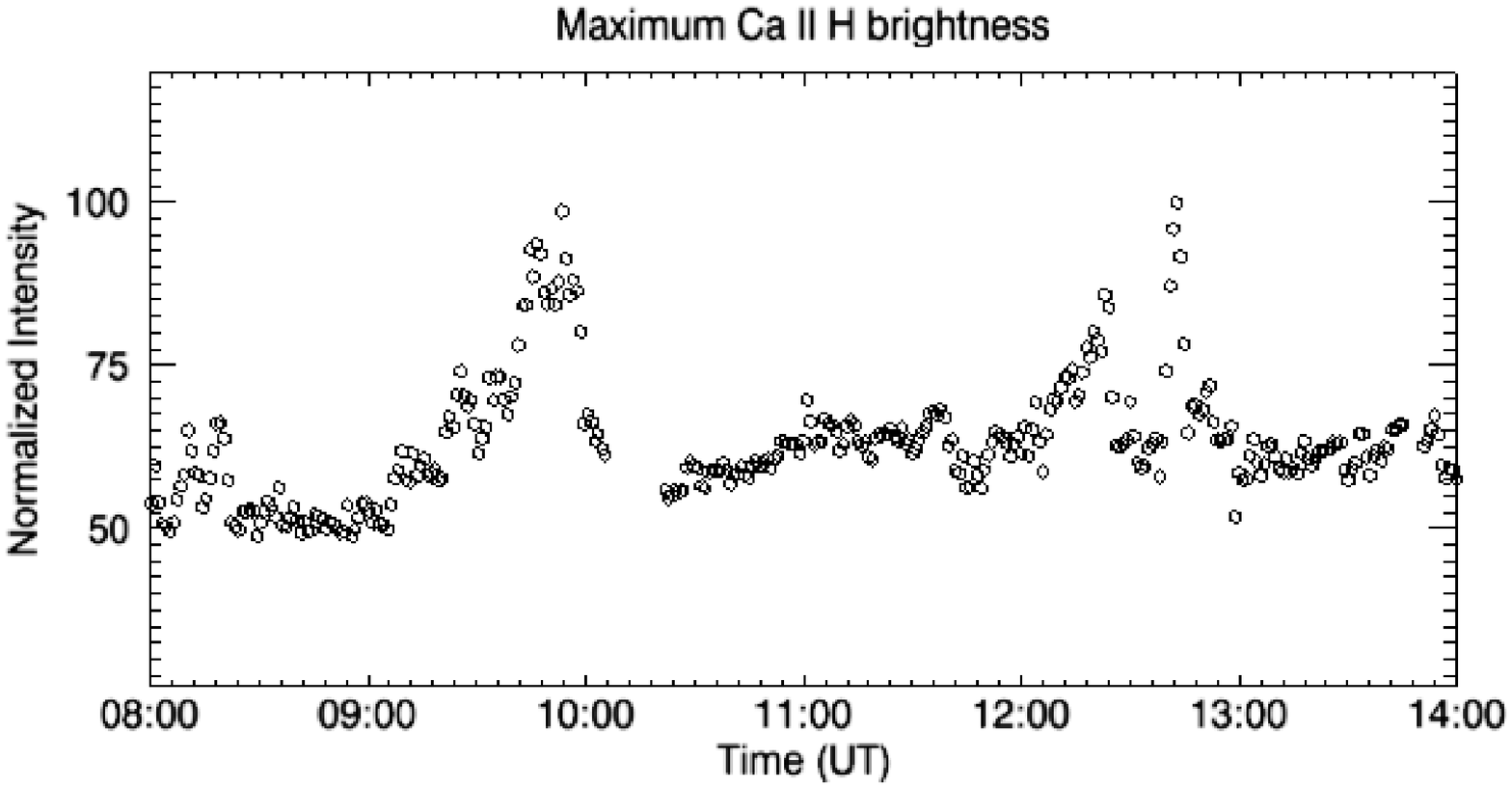}
\plotone{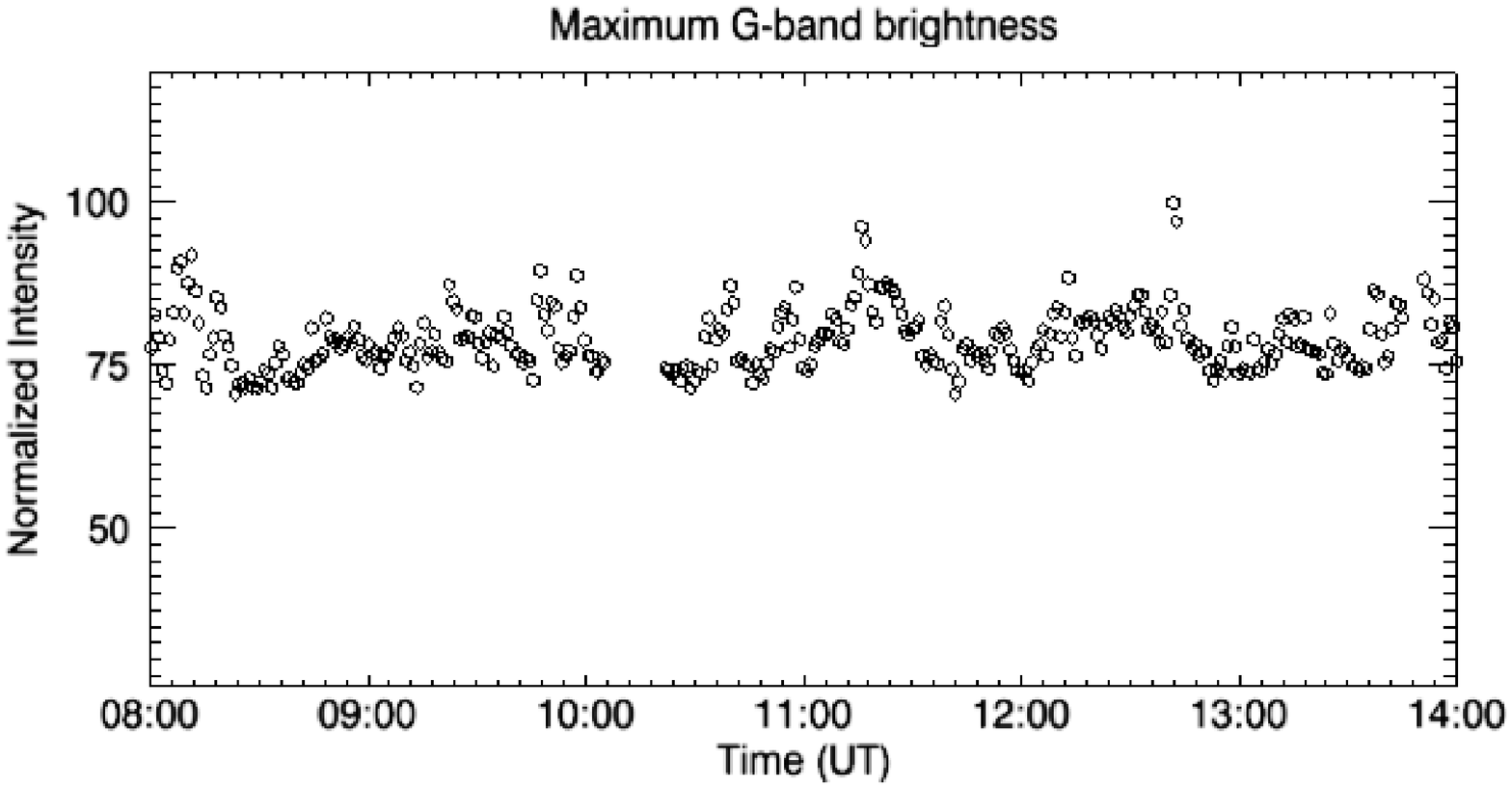}
\caption{\emph{Top:} Evolution of the maximum \ion{Ca}{2} H brightness (normalized DNs). Two main peaks can be observed at 9:50 UT and 12:40 UT. \emph{Bottom}: Same, for G-band intensity. The increases are not correlated with \ion{Ca}{2} H enhancements. \label{fig5}}
\end{figure}

In this area we have detected chromospheric \ion{Ca}{2} H brightness
enhancements with two main peaks during the observations, each one
preceded by a minor peak. As can be seen in Fig.~\ref{fig2}, the
brightenings are associated with the 
positive-polarity footpoint. The duration of each peak is about half
an hour, with an enhancement of $\sim 80\%$ with respect to the
``quiet'' level. The presence of these peaks points to interactions
between the new emerging and the old pre-existing flux systems. We
have computed the average intensity of the four most luminous
pixels within the $8 \times 8 \;\textrm{Mm}^{2}$ FoV for the
\ion{Ca}{2} H and G-band filtergrams, respectively. Figure~\ref{fig5}
shows the trend of brightness in
\ion{Ca}{2} H and G band in normalized units. The different behaviour indicates 
that they are not correlated, as the chromospheric brightness enhancements are 
much more intense than the increase observed in the G band at the same times. 
Thus, the observed \ion{Ca}{2} H enhancements are genuinely due to photons coming 
from the chromosphere, and not to the significant photospheric contribution 
included in the passband of the SOT \ion{Ca}{2} H filter \citep{Carlsson:07}.  

\begin{figure}
\epsscale{1.}
\plotone{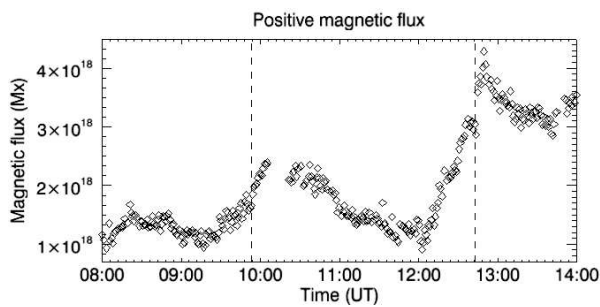}
\caption{Evolution of the positive magnetic flux in the area, brought by the newly emerging bipole. Uncertainties are of the order of the symbol size. The vertical lines indicate the times of the chromospheric brightenings. \label{fig6}}
\end{figure}

We have calculated the positive flux in the $8 \times 8$ Mm$^2$ area
using the \ion{Na}{1} D1 magnetograms, which have a noise level
of approximately $6.75 \times 10^{14} \;\textrm{Mx/pixel}$. The main
contribution to the positive flux comes from the positive polarities
of the emerging region. In Fig.~\ref{fig6} we show the flux evolution
with time: the chromospheric brightness enhancements clearly
correspond to an increase of positive magnetic flux in the upper
photosphere. Interestingly, in both cases the maximum Ca II H
intensities are reached some 30 minutes after the positive flux starts
to increase.

Taking into account the temporal coincidence between the chromospheric
brightenings and the positive flux increase, as well as the spatial
coincidence between the location and morphology of the \ion{Ca}{2} H
brightenings and the emerging bipolar region (Fig.~\ref{fig2}; compare
also the second, third, and fifth rows of Fig.~\ref{fig4}), we conclude
that the localized chromospheric heating is a consequence of the
emergence and subsequent interaction of the positive flux of the new
bipole, which cancels with the negative ambient magnetic field.

\section{Conclusions}

Using \emph{Hinode} filtergrams and spectropolarimetric measurements,
we have studied a small-scale flux emergence event. Two peaks of
chromospheric origin have been detected in the \ion{Ca}{2} H line-core
intensity almost simultaneously to a magnetic flux increase in the
upper photosphere.

We suggest that the chromospheric brightness enhancements may be
indication that two different flux systems undergo magnetic
reconnection: the old flux system belonging to the active region and
the emerging magnetic field. The energy released in the process heats
the chromosphere. The observed \ion{Ca}{2} H brightenings are
associated with a relatively modest amount of emerged magnetic flux
(only $\sim 4 \times 10^{18} \;\textrm{Mx}$ compared with the total negative
flux in the region of $\sim 2.5 \times 10^{19} \;\textrm{Mx}$), which
points to a highly efficient heating mechanism. We conjecture that, 
while the positive flux increases, part of it cancels with the pre-existing 
negative flux, very likely in a process of magnetic reconnection.

Our result suggests that Joule dissipation may be a significant
source of chromospheric heating during the reconnection of an emerging
flux system with a pre-existing magnetic field. This would
confirm the predictions of recent numerical simulations
\citep[e.g.][]{Galsgaard:05}, also at small scales. Moreover, our work
suggests that \ion{Ca}{2} H brightness enhancements can be used as a
valuable diagnostics of flux emergence. Further investigations should
put this result into a firm observational and theoretical basis.

\acknowledgments
Financial support by the European Commission through the SOLAIRE
Network (MTRN-CT-2006-035484) is gratefully acknowledged. This work
has been partly funded by the Spanish Ministerio de Educaci\'on y
Ciencia through projects ESP2006-13030-C06-02, PCI2006-A7-0624, and
Programa de Acceso a Infraestructuras Cient\'ificas y Tecnol\'ogicas
Singulares. 
\emph{Hinode} is a Japanese
mission developed and launched by ISAS/JAXA, with NAOJ as domestic
partner and NASA and STFC (UK) as international partners. It is
operated by these agencies in co-operation with ESA and NSC (Norway).

\onecolumn

\twocolumn

\end{document}